\documentclass[12pt]{article}
\usepackage[margin=2cm]{geometry}
\usepackage{amsmath}
\usepackage{amssymb}
\usepackage{graphicx}
\usepackage{hyperref}

\numberwithin{equation}{section}

\begin{document}
\begin{flushright}
OU-HET 793
\end{flushright}
\vspace{2cm}
\begin{center}
{\LARGE 't Hooft Operators on an Interface and Bubbling D5-Branes}\\
\vspace{2cm}
{\Large Koichi Nagasaki\footnote{nagasaki@het.phys.sci.osaka-u.ac.jp}, Satoshi Yamaguchi\footnote{yamaguch@het.phys.sci.osaka-u.ac.jp}}
\end{center}

\begin{center}
{\it
Department of Physics, Osaka University, Toyonaka 560-0043, Japan
}
\end{center}
\vspace{3cm}
\abstract{We consider a brane configuration consisting of a D5-brane, D1-branes and D3-branes.  According to the AdS/CFT correspondence this system realizes a 't Hooft operator embedded in the interface in the gauge theory side. In the gravity side the near-horizon geometry is $AdS_5\times S^5$.  The D5-brane is treated as a probe in the $AdS_5\times S^5$ and the D1-branes become the gauge flux on the D5-brane.
We examine the condition for preserving appropriate amount of supersymmetry and derive a set of differential equations which is the sufficient and necessary condition.  This supersymmetric configuration shows bubbling behavior.
We try to derive the relation between the probe D5-brane and the Young diagram which labels the corresponding 't Hooft operator. 
We propose the dictionary of the correspondence between the Young diagram and the probe D5-brane configuration. 
}
\thispagestyle{empty}

\newpage

\tableofcontents
\section{Introduction}
Non-local operators play an important role in studying the AdS/CFT correspondence \cite{Maldacena:1997re}. These operators are classified by the dimensionalities of the operators. There must be a string theory counterpart of each operator according to the AdS/CFT correspondence. For example, a Wilson loop operator \cite{Wilson:1974sk} is a 1-dimensional non-local operator and corresponds to a fundamental string \cite{Rey:1998ik,Maldacena:1998im} or a probe D-brane \cite{Rey:1998ik,Drukker:2005kx,Hartnoll:2006hr,Yamaguchi:2006tq,Gomis:2006sb}.
A surface operator is a 2-dimensional non-local operator and corresponds to a D3-brane \cite{Gukov:2006jk,Gomis:2007fi,Drukker:2008wr,Koh:2008kt}. An example of a 3-dimensional non-local operator is ``an interface'' \cite{Sethi:1997zza,Ganor:1997jx,Kapustin:1998pb,Karch:2000gx,DeWolfe:2001pq,Kirsch:2004km,Gomis:2006cu,D'Hoker:2007xy,D'Hoker:2007xz,Gaiotto:2008sa,Gaiotto:2008sd,Gaiotto:2008ak,Nagasaki:2012re,Nagasaki:2011ue}. 

The Wilson loop operator has a bubbling geometry description in the gravity side \cite{Yamaguchi:2006te,Lunin:2006xr,D'Hoker:2007fq,Okuda:2008px} which is an analogue of the bubbling AdS geometry for local operators \cite{Lin:2004nb}.   In this case the total geometry is described as a fiber bundle over 2-dimensional base space.  This base space with the boundary carries the information of the representation of the Wilson loop, or the Young diagram.

On the other hand, the interface is a 3-dimensional non-local operator in the gauge theory. It is known that in the AdS/CFT scenario this operator is introduced by adding a probe D5-brane to the original multiple D3-brane system \cite{Karch:2000gx}.
As a result, some of the D3-branes can end on the D5-brane. The gauge theory realized from this system consists of two gauge theories with different gauge groups divided by a wall --- the interface.

In this paper we consider another type of 1-dimensional non-local operators --- 't Hooft operators \cite{Goddard:1976qe,Kapustin:2005py,Kapustin:2007wm,Saulina:2011qr} on the interface. 
They have magnetic charges while the Wilson operators have electric charges.  
The 't Hooft operators correspond to D1-branes in the string theory. So we can construct the system consisting of D3-branes, a D5-brane and D1-branes so that the supersymmetry is preserved as shown later.
 The expected correspondence is as follows. In the previous bubbling geometry scenario the boundary of the bubbling structure is related to the Young diagram which classifies the Wilson operators. In the same way we expect the boundary of the bubbling D5-brane is related to the Young diagram which classifies the 't Hooft operators. We note that the worldvolume of the probe D5-brane has the bubbling structure while in the bubbling geometries the spacetime geometries have the bubbling structure.
Our goal is to relate the D1-brane system corresponding to the 't Hooft operator to the Young diagram using the probe D5-brane.

We examine the supersymmetry condition of the D1-D5 bound state in the $AdS_5\times S^5$ and obtain a set of differential equations which determines the configuration of the D5-brane worldvolume embedding and the gauge flux on it.
To solve these equations we require the boundary condition. This condition determines the shape of the Young diagram.

The outline of this paper is as follows. In section \ref{BraneConfiguration} we introduce the brane configuration used in our investigation.
In section \ref{ConditionForSUSY}, we study that the structure of the D-branes is restricted by the condition for preserving supersymmetry and derive a set of differential equations. In section \ref{SolvingSUSYCondition} we carefully look at the equations determining the brane structure, and find the independent smaller set of equations.  In section \ref{BoundaryBehaviour} we investigate the boundary of the bound state of the multiple brane system in order to see how to relate the brane system to the Young diagram.  In section \ref{Conclusion} we summarize the result of this paper and propose future works.

\section{Brane configuration}\label{BraneConfiguration}
The AdS/CFT correspondence with a probe D5-brane has been studied in \cite{Karch:2000gx}. 
Let us first briefly review this correspondence.
This system consists of $N$ D3-branes and a D5-brane. 
The D3-branes extend along the directions 0123 in 10-dimensional spacetime and the D5-brane extends 012456 (see table \ref{tab:BraneConfiguration}). The D5-brane does not extend in the direction 3, so D3-branes can end on the D5-brane in this direction. Let $k$ D3-branes out of $N$ end on this D5-branes, and suppose $k\ll N$.  This system can be seen from two different points of view: the gravity side and the gauge theory side.
These two theories are conjectured to be equivalent. 

In the gravity side, these multiple D3-branes warp the spacetime and give rise to $AdS_5\times S^5$ spacetime in the near horizon limit. Meanwhile, the backreaction of the D5-brane is negligible, and therefore the D5-brane is treated as a probe brane. Consequently, this system describes the superstring theory with the probe D5-brane in the $AdS_5\times S^5$. 

In the gauge theory side, the D5-brane is regarded as a wall between gauge theories with different gauge groups $SU(N)$ and $SU(N-k)$ where $N$ is the total number of the D3-branes and $k$ is the number of D3-branes which end on the D5-brane. This wall gives the boundary condition of each gauge theory and is called ``an interface.'' 

In this paper we would like to insert a 't Hooft operator on the interface in the gauge theory. This corresponds to adding D1-branes ending on the D3-branes in string theory.  The total system is then made of $N$ D3-branes, a D5-brane and D1-branes as shown in table \ref{tab:BraneConfiguration}.

\begin{table}[htb]
\begin{center}
\begin{tabular}{|c|c|c|c|c|c|c|c|c|c|c|}
\hline  & 0 & 1 & 2 & 3 & 4 & 5 & 6 & 7 & 8 & 9 \\
\hline D3 & $\circ$ & $\circ$ & $\circ$ & $\circ$ &  &  &  &  &  &  \\
\hline D5 & $\circ$ & $\circ$ & $\circ$ &  & $\circ$ & $\circ$ & $\circ$ &  &  &  \\
\hline D1 & $\circ$ &  &  &  & $\circ$ &  &  &  &  &  \\
\hline \end{tabular} 
\caption{The brane system. In this table ``$\circ$'' denotes the directions along which branes extend.}
\label{tab:BraneConfiguration}
\end{center}
\end{table}

Similar to the previous case, the D3-branes forming the space-time give $AdS_5\times S^5$ geometry, while the D5-brane and the D1-branes are treated as probes. The D1-branes are embedded as a worldvolume flux in the D5-brane and there is a symmetry $U(1)\times U(1)\times SO(3)$ related to the rotations in the directions 12, 56 and 789, respectively. This configuration preserves 1/4 of original supersymmetry in the near-horizon.

\section{Condition for SUSY}\label{ConditionForSUSY}
In this section we study the supersymmetric embedding of the D5-brane. First, we investigate the supersymmetry of the bulk spacetime $AdS_5\times S^5$. Second, a part of the bulk supersymmetry is broken when a D-brane is added. The remaining supersymmetry is analyzed  by the kappa symmetry projection. Finally, this condition gives the restriction to the embedding of the D5-brane in the bulk spacetime.

\subsection{Supersymmetry in the bulk spacetime}
We first consider the supersymmetry of the bulk spacetime $AdS_5\times S^5$.
We concentrate on the $S^2$ part of the $S^5$. The metric is 
\begin{equation}\label{AdS5S5Metric}
ds^2 = \frac{1}{y^2}(-dt^2+dy^2+dr^2+r^2d\psi^2+dx_3^2) + d\theta^2+\sin^2\theta d\phi^2.
\end{equation}
The RR five-form flux $F^{(5)}$ takes non-zero value:
\begin{equation}
F^{(5)}=4({\rm vol}(AdS_5) + {\rm vol}(S^5)),
\end{equation}
where ${\rm vol}(AdS_5)$ and ${\rm vol}(S^5)$ are volume forms of $AdS_5$ and $S^5$ respectively.
Here we use the unit where the radius of the AdS spacetime equals to unity.
The condition for preserving supersymmetry is that the supersymmetry transformations of the fermions are zero. The dilatino condition is trivially satisfied in the above background, while the gravitino condition for the SUSY parameter $\epsilon$ can be written as
\begin{equation}
\nabla_M\epsilon +\frac{i}{2^4}\Gamma^{M_1M_2\cdots M_5}F^{(5)}_{M_1M_2\cdots M_5}\Gamma_M\epsilon =0. \label{gravitino-condition}
\end{equation}
Here the covariant derivative is defined as
\begin{equation}
\nabla_M = \partial_M +\frac{1}{4}\Omega_M{}^{AB}\Gamma_{AB}, 
\end{equation}
where $\Omega_{M}{}^{AB}$ are the spin connections which are related to the vielbein $E^M,\: M=0,1,\cdots 9$ of the metric \eqref{AdS5S5Metric} as
\begin{equation}
dE^A=-\Omega^A_{\:\:B}E^B, \qquad
\Omega^{AB}=-\Omega^{BA}, \qquad
\Omega^{AB}=\Omega_M{}^{AB}E^M.
\end{equation}
We choose the vielbein as
\begin{align}
& E^0=\frac{dt}{y},\:\: E^1=\frac{dr}{y},\:\: E^2=\frac{rd\psi}{y},\:\: E^3=\frac{dx_3}{y},\:\: E^4=\frac{dy}{y},\nonumber\\
& E^5=d\theta,\:\: E^6=\sin\theta d\phi.
\end{align}
For the detailed calculation of eq.~\eqref{gravitino-condition}, see appendix \ref{BulkSUSY}.  Here we only show the result.
The bulk space preserves the SUSY generated by the parameter
\begin{equation}\label{SUSYparameter}
\epsilon=e^{-\frac{\theta}{2}\gamma\Gamma_{45}}e^{\frac{\phi}{2}\Gamma_{56}}e^{-\frac{1}{2}\ln y\cdot \gamma}
e^{r\frac{1+\gamma}{2}\Gamma_{14}}
e^{x_3\frac{1+\gamma}{2}\Gamma_{34}}
e^{t\frac{1+\gamma}{2}\Gamma_{04}}
e^{\frac{\psi}{2}\Gamma_{12}}\epsilon_0,
\end{equation}
where $\gamma=-i\Gamma_{0123}$ and $\epsilon_0$ is a constant spinor as we calculated in appendix \ref{BulkSUSY}.
\subsection{Ansatz for D5-brane}\label{OurAnsatz}
We consider a bound state of a D5-brane and D1-branes in the $AdS_5\times S^5$ spacetime. The D1-branes are realized as the worldvolume gauge flux on the D5-brane. Thus we consider a probe D5-brane with the worldvolume gauge flux.
We define the worldvolume coordinates of the D5-brane as $(t, y, \psi, \phi, u_1, u_2)$ where the coordinates $(t, y, \psi, \phi)$ are identified with the coordinates of the bulk spacetime. According to the symmetry $U(1)^2\times SO(3)$ we put the ansatz on the embedding as:
\begin{eqnarray}\label{Ansatz}
r=ys(u),\:\: x_3=yz(u),\:\: \theta=\theta(u),
\end{eqnarray}
where $s(u)$, $z(u)$ and $\theta(u)$ are unknown functions of coordinates $u^i$, $i=1,2$. Since $(u^1, u^2)$ are not fixed yet, there remains the general coordinate transformation symmetry of $(u^1,u^2)$. Some of the D3-branes end on the D5-brane. Thus the  ansatz for the worldvolume gauge flux is written as
\begin{equation}\label{Flux}
\mathcal{F} = dP \wedge d\psi + dQ \wedge d\phi,
\end{equation}
where potentials $P$ and $Q$ are functions of $u$. Then we have unknown functions of $u$
\begin{equation}\label{UnknownFunctions}
s(u), \:\:  z(u), \:\:  \theta(u), \:\:  P(u),\:\: Q(u).
\end{equation}
Our goal is to determine these functions.
\subsection{Condition for SUSY}
In this subsection we try to obtain the condition for preserving supersymmetry. 
When a D$p$-brane exists, a part of the original supersymmetry is broken. The remaining supersymmetry parameters are spinors of the form \eqref{SUSYparameter} which satisfy the relation
\cite{Cederwall:1996pv,Aganagic:1996pe,Cederwall:1996ri,Bergshoeff:1996tu,Aganagic:1996nn,Bergshoeff:1997kr,Skenderis:2002vf}
\begin{equation}\label{KappaProj}
\Gamma\epsilon=\epsilon.
\end{equation}
This is called  ``the kappa symmetry projection'' where the operator $\Gamma$ is determined for a D$p$-brane as
\begin{align}
& d^{p+1}\xi \cdot\Gamma := \left(-e^{-\Phi}(-\det(G_{\rm ind}+\mathcal{F}))^{-1/2}e^{\mathcal{F}}\chi\right)\Big|_{(p+1)-{\rm form}},\\
& \chi := \sum_{n} \frac{1}{(2n)!} \hat{E}^{a_{s}} \cdots \hat{E}^{a_1} \Gamma_{a_1\cdots a_s} K^n(-i),
\end{align}
where $\xi^i$, $i=0,\cdots, p$, are worldvolume coordinates, $\Phi$ is the dilaton, $G_{\rm ind}$ is the induced metric of the D$p$-brane and $\hat{E}^A$ is the pullback of $E^A$ defined as $\hat{E}^A:= E^A_M \frac{\partial X^M}{\partial \xi^i} d\xi^i$. We calculated examples for a D5-brane and for a D1-brane in appendix \ref{GammaForD5D1} and we use the relations obtained in these examples in the following calculation.

We calculate the kappa symmetry projection operator $\Gamma$ defined above under our ansatz given in section \ref{OurAnsatz}.  Here we only show the result
\begin{equation}\label{GammaProj}
\Gamma = \frac{1}{W} \Big\{
s\sin\theta \:\mathcal{A} \Gamma_{62}K(-i)\Gamma_{04}
+\sin\theta \:\mathcal{B} (-i)\Gamma_{60}
-s\:\mathcal{C}(-i)\Gamma_{20}
+\mathcal{D} K(-i) \Gamma_{04}
\Big\}.
\end{equation}
For the detailed calculation, see appendix \ref{CalcGamma}.
Here we defined a $y$ independent function $W$ as
\begin{equation}\label{DBIfactor}
W:=y^2\sqrt{-\det(G_{\rm ind}+\mathcal{F})}.
\end{equation}
The induced metric for the D5-brane is 
\begin{align}
ds^2=-\frac{1}{y^2}dt^2+s^2d\psi^2+\sin^2\theta d\phi^2 + \frac{\beta}{y^2}dy^2
	+h_{ij} du^i du^j +\frac{\partial_a\beta}{y}du^ady,\\
\beta:=1+s^2+z^2, \:\:\: h_{ij}:=\sum_{\lambda=s,z,\theta}\partial_i \lambda \partial_j \lambda\nonumber.
\end{align}
In the expression \eqref{GammaProj}, $\mathcal{A},\mathcal{B},\mathcal{C},\mathcal{D}$ are the following matrices.
\begin{subequations}
\renewcommand{\theequation}{\theparentequation--\roman{equation}}
\begin{align}
&\mathcal{A} := -\{s,z\}\Gamma_{13}-\{s,\theta\}\Gamma_{15} -\{z,\theta\}\Gamma_{35} + s^2\{\frac{z}{s},\theta\}\Gamma_{1345},\\ 
&\mathcal{B} := -\{P,\frac{z}{s}\}\Gamma_{13}+\{P,s\}\Gamma_{14}+ \{P,z\}\Gamma_{34} -s\{P,\theta\}\Gamma_{15} -z\{P,\theta\}\Gamma_{35}-\{P,\theta\}\Gamma_{45},\\
&\mathcal{C} := -\{Q,\frac{z}{s}\}\Gamma_{13}+\{Q,s\}\Gamma_{14}+ \{Q,z\}\Gamma_{34} -s\{Q,\theta\}\Gamma_{15} -z\{Q,\theta\}\Gamma_{35}-\{Q,\theta\}\Gamma_{45},\\
&\mathcal{D} := -\{P,Q\} (1+s\Gamma_{14}+z\Gamma_{34}),
\end{align}
\end{subequations}
where $\mathcal{C}$ is obtained from $\mathcal{B}$ by replacing all $P$'s by $Q$'s. We use the notation of ``Poisson bracket''
\begin{equation}\label{Bracket}
\{A,B\} := \epsilon^{ab}\frac{\partial A}{\partial u^a}\frac{\partial B}{\partial u^b}
=\frac{\partial A}{\partial u^1}\frac{\partial B}{\partial u^2} - \frac{\partial A}{\partial u^2}\frac{\partial B}{\partial u^1}.
\end{equation}

Under our ansatz the parameter $\epsilon$ in eq.~\eqref{SUSYparameter} is decomposed by the dependence of $y$ and $t$:
\begin{align}
\epsilon 
&= e^{-\frac{\theta}{2}\gamma\Gamma_{45}}e^{\frac{\phi}{2}\Gamma_{56}}e^{-\frac{1}{2}\ln y\cdot \gamma}
	e^{r\frac{1+\gamma}{2}\Gamma_{14}}
	e^{x_3\frac{1+\gamma}{2}\Gamma_{34}}
	e^{t\frac{1+\gamma}{2}\Gamma_{04}}
	e^{\frac{\psi}{2}\Gamma_{12}}\epsilon_0\nonumber\\
&= e^{-\frac{\theta}{2}\gamma\Gamma_{45}}e^{-\frac{1}{2}\ln y\cdot \gamma}
	\left(1+ ys\frac{1+\gamma}{2}\Gamma_{14}\right)
	\left(1+ yz\frac{1+\gamma}{2}\Gamma_{34}\right)
	\left(1+ t\frac{1+\gamma}{2}\Gamma_{04}\right)
	\xi\nonumber\\
&= e^{-\frac{\theta}{2}\gamma\Gamma_{45}}e^{-\frac{1}{2}\ln y\cdot \gamma}
	\left(\xi+ ys\Gamma_{14}\xi_{-}
		+ yz\Gamma_{34}\xi_{-}
		+ t\Gamma_{04}\xi_{-}\right)\nonumber\\
&= e^{-\frac{\theta}{2}\gamma\Gamma_{45}}
	\left(\frac{1}{\sqrt{y}}\xi_{+}+\sqrt{y}\xi_{-}
		+ \frac{1}{\sqrt{y}}(ys\Gamma_{14}\xi_{-}
		+ yz\Gamma_{34}\xi_{-}
		+ t\Gamma_{04}\xi_{-})\right)\nonumber\\
&=: \sqrt{y} \epsilon_1 + \frac{1}{\sqrt{y}} \epsilon_2 + \frac{t}{\sqrt{y}} \epsilon_3,
\end{align}
where we define $\xi := e^{\frac{\phi}{2}\Gamma_{56}}e^{\frac{\psi}{2}\Gamma_{12}}\epsilon_0$ in the second line and $\xi_{\pm}:=\frac{1\pm \gamma}{2}\xi$.  The explicit forms of $\epsilon_1,\epsilon_2,\epsilon_3$ are written as
\begin{subequations}
\renewcommand{\theequation}{\theparentequation--\roman{equation}}
\begin{align}
&\epsilon_1 = e^{-\frac{\theta}{2}\gamma\Gamma_{45}} (1+s\Gamma_{14}+z\Gamma_{34})\xi_{-},\\
&\epsilon_2 = e^{-\frac{\theta}{2}\gamma\Gamma_{45}} \xi_{+}, \label{epsilon2}\\
&\epsilon_3 = e^{-\frac{\theta}{2}\gamma\Gamma_{45}} \Gamma_{04}\xi_{-}.
\end{align}
\end{subequations}
Since the kappa symmetry operator of eq.~\eqref{GammaProj} is independent of $y$ and $t$, we can impose the projection condition \eqref{KappaProj} for each $\epsilon_i$ :
\begin{equation}\label{Projection123}
\Gamma\epsilon_i = \epsilon_i, \:\:\: i= 1,2,3.
\end{equation}

The kappa symmetry projections for the D5-brane and the D1-brane give the conditions \eqref{KappaConditionD5} and \eqref{KappaConditionD1}, respectively, which are obtained in appendix \ref{GammaForD5D1}.
\begin{align}
& {\rm D5 \:\: condition} \Leftrightarrow (K\Gamma_{3456}+\gamma)\xi=0 \label{KappaProjD5},\\
& {\rm D1 \:\: condition} \Leftrightarrow (iK\Gamma_{04}-1)\xi=0 \label{KappaProjD1}.
\end{align}
We want to obtain the condition for the functions \eqref{UnknownFunctions} such that all spinors restricted by the equations \eqref{KappaProjD5} and \eqref{KappaProjD1} satisfy the projection condition \eqref{Projection123}.
The condition \eqref{Projection123} is equivalent to 
\begin{align}\label{ABCDepsilon}
e^{\frac{\theta}{2}\gamma\Gamma_{45}}\Big\{
s\sin\theta \:\mathcal{A} \Gamma_{62}K(-i)\Gamma_{04}
+\sin\theta \:\mathcal{B} (-i)\Gamma_{60}
-s\:\mathcal{C}(-i)\Gamma_{20}
+\mathcal{D} K(-i) \Gamma_{04}-W
\Big\}\epsilon_i=0, \nonumber\\ i=1,2,3.
\end{align}
For $\epsilon_2$ \eqref{epsilon2}, 
\begin{equation}\label{ABCDepsilon2}
\eqref{ABCDepsilon}\Leftrightarrow
\Big\{
s\sin\theta \:\mathcal{A} \cdot\Gamma_{51}
+\sin\theta \:\mathcal{B} \Gamma_{53}e^{\theta\Gamma_{45}}
-s\:\mathcal{C}\cdot\Gamma_{31}
+\mathcal{D} e^{\theta\Gamma_{45}}-W
\Big\}\xi_{+}=0,
\end{equation}
where we used relations obtained from \eqref{KappaProjD5}, \eqref{KappaProjD1} and $\gamma\xi_{\pm}=-i\Gamma_{0123}\xi_{\pm}=\pm\xi_{\pm}$,
\begin{subequations}
\renewcommand{\theequation}{\theparentequation--\roman{equation}}
\begin{align}
& \Gamma_{62}\xi_{\pm}=\Gamma_{51}\xi_{\pm},\\
& \Gamma_{60}\xi_{\pm}=\pm i\Gamma_{53}\xi_{\pm},\\
& \Gamma_{20}\xi_{\pm}=\pm i\Gamma_{31}\xi_{\pm}.
\end{align}
\end{subequations}
The left hand side of \eqref{ABCDepsilon2} can be written only by using $\Gamma_1,\Gamma_3,\Gamma_4,\Gamma_5$ and ${\bf 1}$ (identity matrix) and their products. Each coefficient of independent matrices gives the conditions:
\begin{subequations}
\renewcommand{\theequation}{\theparentequation--\roman{equation}}
\begin{align}
& s\{s,\cos\theta\}-\sin\theta\{P,z\sin\theta\}+s^3\{Q,\frac{z}{s}\}-\cos\theta\{P,Q\}-W=0 \label{Fromep23b}, \\
& s\{z,\theta\} -\{P,s\sin\theta\}=0, \\
& s\sin^2\theta\{P,\frac{z}{s}\}+\{Q,z\}+\cos\theta\{P,Q\}=0, \\
& s^2\sin\theta\cos\theta\{P,\frac{z}{s}\}+sz\{Q,\theta\}-s\sin\theta\{P,Q\}=0\label{gamma155},\\
& s^3\{\frac{z}{s},\cos\theta\}+\frac{1}{2}\{P,\cos^2\theta\}-s\{Q,s\}+z\cos\theta\{P,Q\}=0, \\
& \{P,z\cos\theta\}-\{P,Q\}=0, \label{gamma15}\\
& \sin\theta\{P,s\cos\theta\}-s\{Q,\theta\}=0 \label{Fromep23e}.
\end{align}
\end{subequations}
In this equations \eqref{gamma155} is not independent and can be lead from \eqref{gamma15} and \eqref{Fromep23e}.
For $\epsilon_3$, a similar calculation gives the same conditions. 
For $\epsilon_1$, the calculation is a bit complicated, but we can do it in the same way.
\begin{subequations}
\renewcommand{\theequation}{\theparentequation--\roman{equation}}
\begin{align}
& -\frac{s^4}{2}\{\frac{\beta}{s^2},\cos\theta\} +\frac{z^3\sin^2\theta}{2}\{P,\frac{\beta}{z^2}\}-\beta\cos\theta\{P,Q\}-W=0 \label{Fromep1b}, \\
& \frac{sz^3}{2}\{\frac{\beta}{z^2}, \cos\theta\} +\frac{s^3\sin^2\theta}{2}\{P,\frac{\beta}{s^2}\} +\frac{s}{2}\{Q,\beta\}=0 \label{}, \\
& \frac{1}{2}\{\beta,\cos\theta\} - \frac{z^3}{2}\{Q,\frac{\beta}{z^2}\} -W=0, \\
& \frac{1}{2}\{P,\beta\sin^2\theta\}-\frac{s^4}{2}\{Q,\frac{\beta}{s^2}\}+zW=0, \\
& \cos\theta\{s^2,z\} + \{P,\beta\cos^2\theta\}=0, \\
& \frac{1}{4}\{s^2,z^2\} + \frac{z^3\cos\theta}{2}\{P,\frac{\beta}{z^2}\} +\beta\{P,Q\}=0, \\
& s\sin\theta\{s,z\} + \frac{s^2\sin\theta\cos\theta}{2}\{P,\frac{\beta}{s^2}\} +\beta\{Q,\theta\}=0 \label{Fromep1e}.
\end{align}
\end{subequations}
Consequently, we obtain the 14 equations \eqref{Fromep23b}-\eqref{Fromep23e} and \eqref{Fromep1b}-\eqref{Fromep1e}. We find independent set of these equations in the next section.
\section{Solving SUSY condition}\label{SolvingSUSYCondition}
One can check the last seven equations, \eqref{Fromep1b}-\eqref{Fromep1e}, are derived from eqs.~\eqref{Fromep23b}-\eqref{Fromep23e}. 
So we only have to consider eqs.~\eqref{Fromep23b}-\eqref{Fromep23e} which are rewritten as
\begin{subequations}
\renewcommand{\theequation}{\theparentequation--\roman{equation}}
\begin{align}
& \{s,z\}=-\frac{1}{s\cos\theta}\{P,\beta\cos^2\theta\} \label{Eqsz},\\
& \{s,\theta\}  = -\frac{1}{s}\{P,z\sin\theta\} +\frac{s^2}{\sin\theta}\{Q,\frac{z}{s}\} -\frac{1}{s}\cot\theta\{P,Q\} -\frac{1}{s\sin\theta}W \label{EqsTheta}, \\
& \{z,\theta\}  = \frac{1}{s}\{P,s\sin\theta\} \label{EqzTheta},\\
& \{Q,s\}	=  s^2\{\frac{z}{s},\cos\theta\}+ \frac{1}{2s}\{P,\cos^2\theta\}+\frac{1}{2s}\{P,z^2\cos^2\theta\} \label{EqQs},\\
& \{Q,z\}	= -s\sin^2\theta\{P,\frac{z}{s}\}-\cos\theta\{P,z\cos\theta\} \label{EqQz},\\
& \{Q,\theta\}	= \frac{\sin\theta}{s}\{P,s\cos\theta\} \label{QTheta},\\
& \{P,Q\}	= \{P,z\cos\theta\} \label{EqPQ}.
\end{align}
\end{subequations}
By the definition of the Poisson bracket \eqref{Bracket}, the bracket can be rewritten in terms of differential forms as
\begin{equation}
\{A,B\}du^1\wedge du^2=\partial_i A\partial_j B\epsilon^{ij}du^1du^2=dA\wedge dB=d(A\wedge dB).
\end{equation}
Then eqs.~\eqref{Eqsz}-\eqref{EqPQ} are expressed in terms of differential forms as follows.
\begin{subequations}
\renewcommand{\theequation}{\theparentequation--\roman{equation}}
\begin{align}
& d(\sqrt{\beta}(dz-\cos\theta dP))=0 \tag{\ref{Eqsz}$'$}\label{DiffEq1},\\
& sds\wedge d(\cos\theta)-\sin\theta dP\wedge d(z\cos\theta)+s^3dQ\wedge d(\frac{z}{s})-\cos\theta dP\wedge dQ-Wdu^1\wedge du^2=0 \tag{\ref{EqsTheta}$'$}\label{DiffEq2},\\
& sdz\wedge d(\cos\theta)+\sin\theta dP\wedge d(s\sin\theta)=0 \tag{\ref{EqzTheta}$'$}\label{DiffEq3},\\
& d(P+Q)\wedge \frac{dz}{z}-\frac{\sin^2\theta}{s}dP\wedge ds-\sin\theta dP\wedge d(\sin\theta)=0 \tag{\ref{EqQs}$'$}\label{DiffEq4},\\
& sdQ\wedge ds-\frac{1}{2}dP\wedge d((z^2+1)\cos^2\theta)-s^3d(\frac{z}{s})\wedge d(\cos\theta)=0 \tag{\ref{EqQz}$'$}\label{DiffEq5},\\
& sdQ\wedge d(\cos\theta)+\sin^2\theta dP\wedge d(s\cos\theta)=0 \tag{\ref{QTheta}$'$}\label{DiffEq6},\\
& d(dP(Q-z\cos\theta))=0 \tag{\ref{EqPQ}$'$}\label{DiffEq7}.
\end{align}
\end{subequations}
Since eq.~\eqref{DiffEq6} can be written as a total derivative, it is expressed as the derivative of a appropriate function $\omega$ according to Poincar\'e's lemma:
\begin{align}
& d(-(Q+\sin^2\theta P)\frac{d\theta}{\sin\theta\cos\theta}+P\frac{ds}{s})=0
 \quad \Leftrightarrow \quad  -(Q+\sin^2\theta P)\frac{d\theta}{\sin\theta\cos\theta}+P\frac{ds}{s}=d\omega.
\end{align}
Eqs.~\eqref{DiffEq3}, \eqref{DiffEq7} lead to the relation
\begin{equation}
z\cos\theta=P+Q.
\end{equation}
Furthermore, eq.~\eqref{DiffEq4} and eq.~\eqref{DiffEq5} are equivalent to eq.~\eqref{DiffEq6} and eq.~\eqref{DiffEq1}, respectively.  We also substitute the explicit form of $W$ into eq.~\eqref{EqsTheta}.
Then our equations are simplified as follows.
\begin{subequations}
\renewcommand{\theequation}{\theparentequation--\roman{equation}}
\begin{align}
& d(\sqrt{\beta}(dz-\cos\theta dP))=0,\label{Latestform1}\\
& d\left(-(Q+\sin^2\theta P)\frac{d\theta}{\sin\theta\cos\theta}+P\frac{ds}{s}\right)=0,\label{Latestform2}\\
& z\cos\theta=P+Q,\label{Latestform3}\\
& \left(\frac{s^2}{\cos^2\theta}+1\right)\{P,\cos\theta\}^2 +\frac{s^2}{\cos^2\theta}\{Q,\theta\}^2
  +\frac{s}{\cos\theta}\{s,\cos\theta\}\bigg(\{P,Q\}-z\{P,\cos\theta\}\bigg)\nonumber\\
&\qquad\qquad  +z\{P,\cos\theta\}\{P,Q\}+2\frac{s^2}{\cos^2\theta}\{P,\cos\theta\}\{Q,\cos\theta\}=0\label{Latestform4}.
\end{align}
\end{subequations}
This is one of the main results of this paper.
\subsection{Special case}\label{SpecialCase}
Let us check the consistency of these equations in the well known case \cite{Karch:2000gx} where
\begin{equation}
P=0,\:\: Q=\kappa\cos\theta,\:\: z=\kappa.
\end{equation}
We can easily check that this configuration satisfies eqs.~\eqref{Latestform1}-\eqref{Latestform4}.

This configuration contains no D1-brane and corresponds 
to the 't Hooft operator with the trivial Young diagram.

\section{Boundary behavior}\label{BoundaryBehaviour}
We have to give boundary conditions to solve the equations \eqref{Latestform1}-\eqref{Latestform4}. The boundary of the $u$-plane (the base 2-dimensional space coordinated by $(u^1,u^2)$) is given by $s=0$ or $\sin \theta=0$. 
The boundary condition is not arbitrary and it contains the detailed information of the associated operators in the gauge theory as in \cite{Lin:2004nb,Yamaguchi:2006te,Lunin:2006xr,D'Hoker:2007fq,Okuda:2008px}.
We explain the relation between the boundary behavior of our system and Young diagrams which label the 't Hooft operators.

\begin{figure}[h]
\begin{center}
\includegraphics[width=8cm]{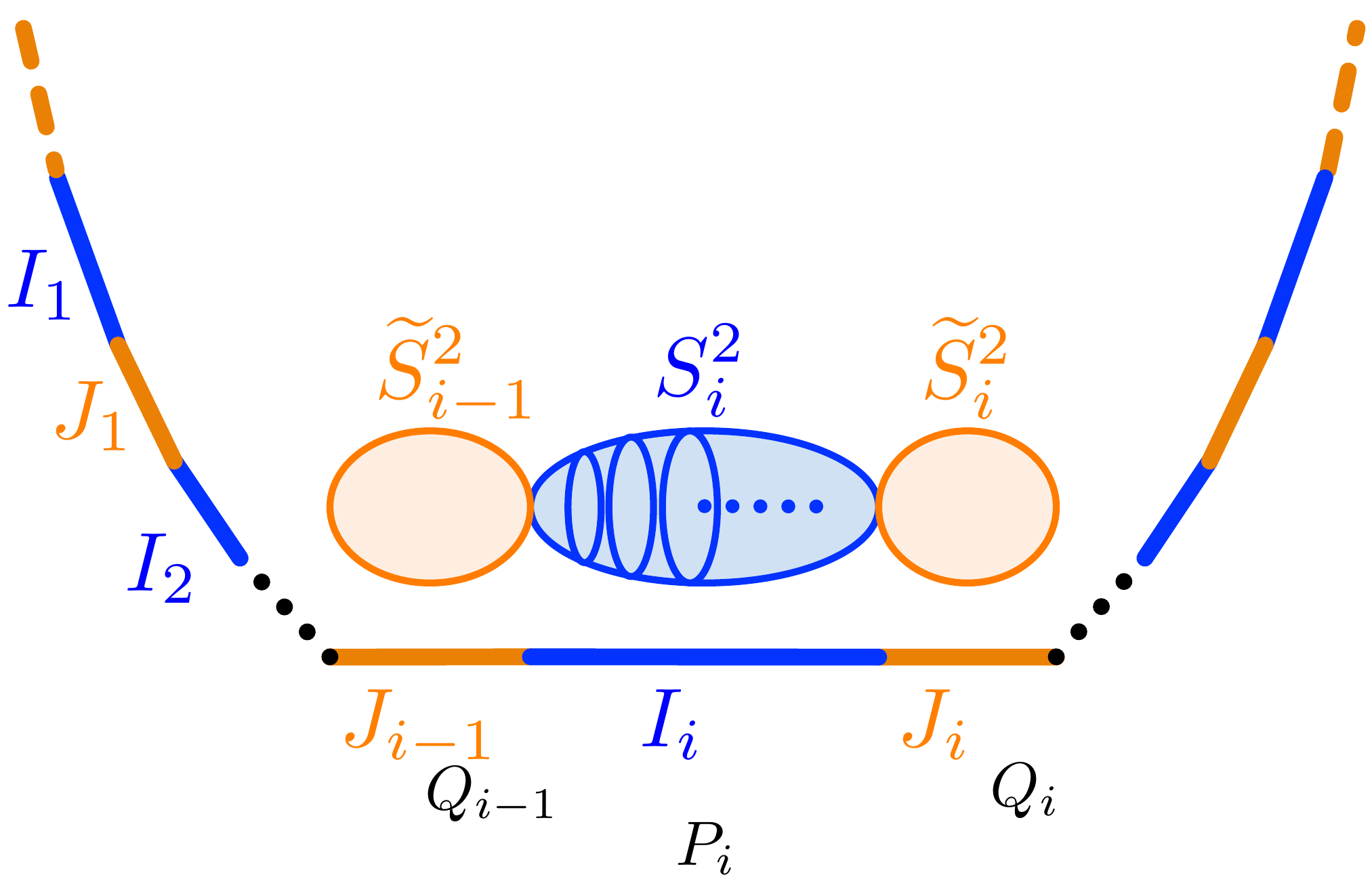}
\end{center}
\caption{The boundary line and 2-spheres composed of $\psi$ and $\phi$-cycles.}
\label{fig:Boundary}
\end{figure}

The structure of the D5-brane worldvolume is a fiber bundle over the $u$-plane with the fiber $S^1\times S^1$ coordinated by $\phi$ and $\psi$. 
Each point of the boundary is distinguished by whether $s=0$ or $\sin\theta=0$ and the boundary is divided into segments as shown in figure \ref{fig:Boundary}. Let $I_i,\ i=1,\dots,\ell$ denote the $i$-th $s=0$ segment and $J_j,\ j=1,\dots,\ell-1$ denote the $j$-th $\sin\theta=0$ segment.
The pullback $dP|_{I_i}$ vanishes and $P$ is a constant $P_i$ on $I_i$ for smoothness since $d\psi$ is singular at $I_i$ and $dP\wedge d\psi$ must vanish.  
The pullback $dQ|_{J_j}$ also vanishes and $Q$ is a constant $Q_j$ on $J_j$ in the same way.
Thus 
the gauge flux reduces to $\mathcal{F}=dQ\wedge d\phi$ at $I_i$ and $\mathcal{F}=dP\wedge d\psi$ at $J_j$. 
At each internal point on $I_i$ the fiber reduces to $S^1$ coordinated by $\phi$ and at both end points of $I_i$ the radius of this $S^1$ fiber vanishes.  Therefore these $S^1$ fibers make a non-contractible $S^2$ cycle denoted by $S^2_i$. There is also a non-contractible $S^2$ cycle (denoted by $\widetilde{S}^2_j$) on $J_j$ in the same way.

The charge is defined as the integration of the flux on each non-contractible $S^2$ and we define these quantities as
\begin{align}
n_i := \frac{\sqrt{\lambda}}{(2\pi)^2} \int_{S_i^2} dQ\wedge d\phi 
= \frac{\sqrt{\lambda}}{2\pi} \int_{I_i}dQ
= \frac{\sqrt{\lambda}}{2\pi} (Q_i-Q_{i-1})
 \label{Defni},\\
m_j := \frac{\sqrt{\lambda}}{(2\pi)^2} \int_{\widetilde{S}_j^2} dP\wedge d\psi 
= \frac{\sqrt{\lambda}}{2\pi} \int_{J_j}dP 
= \frac{\sqrt{\lambda}}{2\pi} (P_{j+1}-P_{j}) 
\label{Defmi}.
\end{align}
Here $Q_0$ is defined as the value of $Q$ on the first $\theta=0$ half line $J_0$.
The normalization is determined so that $n_i$ and $m_j$ are integers as follows.
In a general D5-brane with worldvolume flux the number of the D3-branes and the number of the D1-branes are calculated by the integration of the gauge flux as seen from the Wess-Zumino term of the D5-brane action.
\begin{align}
& (\text{number of D3-branes}) =\frac{T_5}{T_3}\int_{\mathcal{M}_2} \mathcal{F}
	=\frac{1}{(2\pi)^2\alpha'}\int_{\mathcal{M}_2} \mathcal{F},\\
& (\text{number of D1-branes}) = \frac{T_5}{T_1}\int_{\mathcal{M}_4}\frac12 \mathcal{F}\wedge\mathcal{F}
	=\frac{1}{32\pi^4 \alpha'^2}\int_{\mathcal{M}_4} \mathcal{F}\wedge\mathcal{F},
\end{align}
where the integral over $\mathcal{M}_2$ or $\mathcal{M}_4$ denotes the integral over the perpendicular directions to D3-branes or D1-branes on the D5-brane worldvolume. 
We also use the D$p$-brane tension $T_p$
\begin{equation}
T_p=\frac{1}{(2\pi)^p\alpha'^{(p+1)/2}g_s},
\end{equation}
and $\alpha'=1/\sqrt{\lambda}$ in our unit.  Here $g_s$ is the string coupling constant.

\begin{figure}[h]
\begin{center}
\includegraphics[width=8cm]{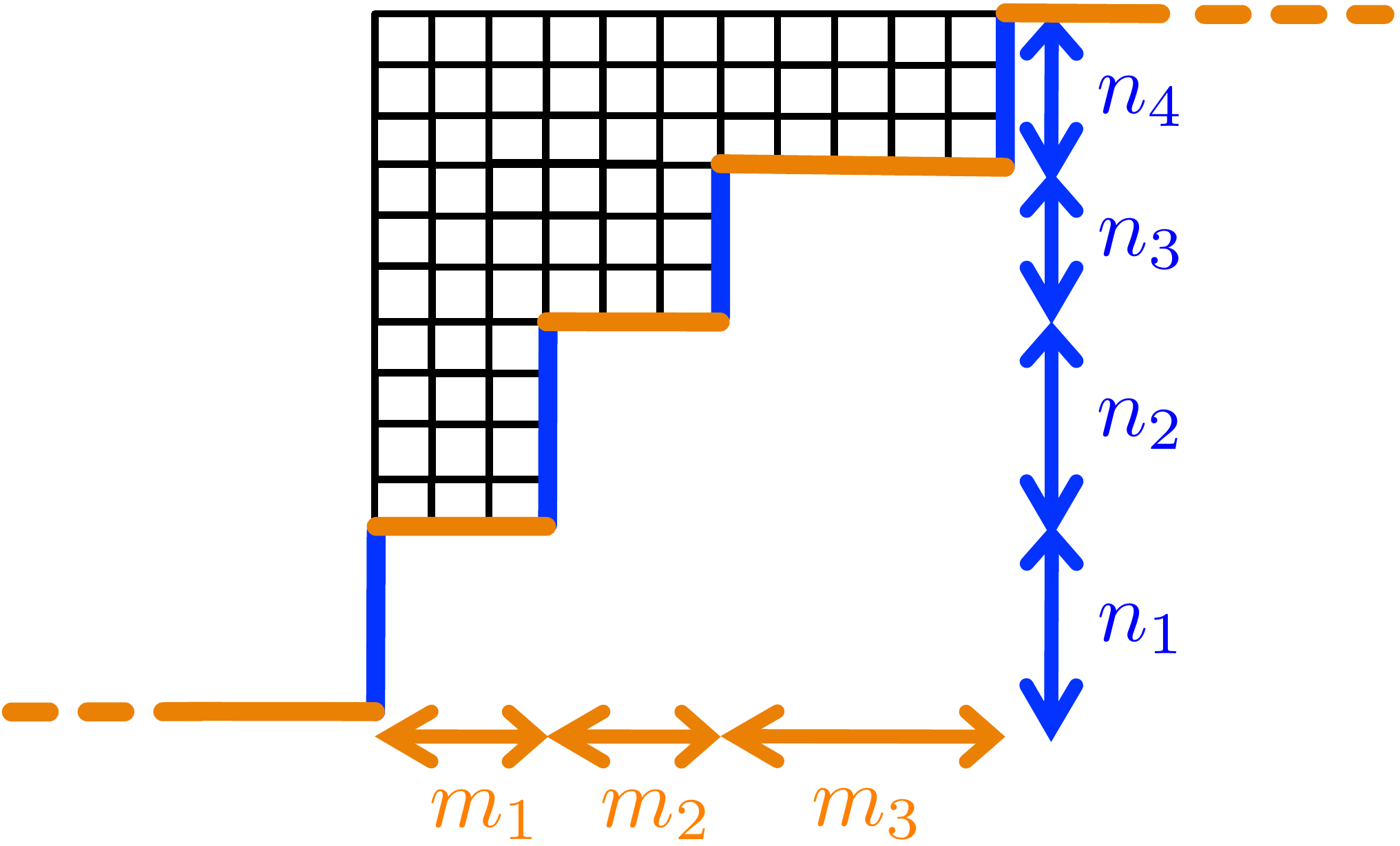}
\end{center}
\caption{The relation between a deformed boundary line and the Young diagram.}
\label{fig:YoungD}
\end{figure}

Since the quantities $n_i$ and $m_j$ are integers, these can be related to the number of boxes in the Young diagram as follows.  First we deform the boundary as stepwise by bending it at the edges of each segment. After that deformation this boundary line can be interpreted as the right down edge of the Young diagram as shown in figure \ref{fig:YoungD}.
The integers $n_i$ and $m_j$ correspond to each length of the edge of the Young diagram.
 
\begin{figure}[h]
\begin{center}
\includegraphics[width=8cm]{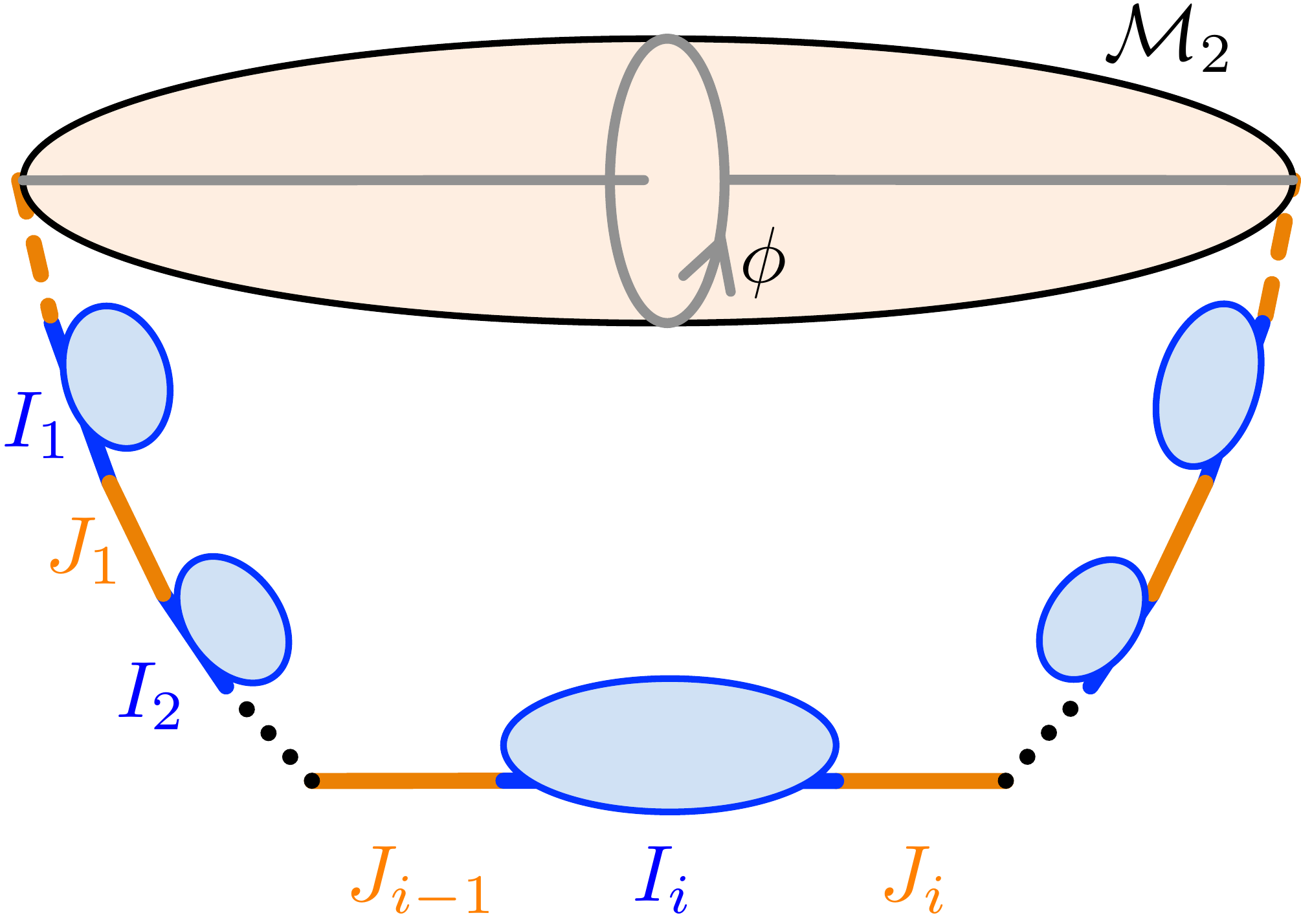}
\end{center}
\caption{$\mathcal{M}_2$ is a 2-dimensional manifold located at sufficiently far. It can be deformed into 2-spheres located in the boundary without changing the value of the integral.}
\label{fig:BoundaryM2}
\end{figure}

Let us consider the relation between the number of branes and the Young diagram for a consistency check. 
The number of the D3-branes ending on the D5-brane, denoted by $k$, is related to the vertical length of the Young diagram as follows.
\begin{align}
k &= \frac{\lambda}{4\pi^2} \int_{\mathcal{M}_2} \mathcal{F} \nonumber\\
   &= \frac{\lambda}{4\pi^2} \sum_{i}\int_{S_i^2} dQ \wedge d\phi \nonumber\\ 
   &= \frac{\lambda}{2\pi} \sum_{i}\int_{I_i} dQ \nonumber\\
   &= \sum_{i}n_i,\label{CalcD3number}
\end{align}
where $\mathcal{M}_2$ is a 2-cycle shown in figure \ref{fig:BoundaryM2}.

On the other hand, the number of the D1-branes $k'$ can be interpreted as the total number of boxes in the Young diagram which characterize the boundary condition as expected. 
This relation is derived as follows.
\begin{align}
k' 	&= \frac{\lambda}{32\pi^4}\int_{\mathcal{M}_4}2\:\: dP\wedge d\psi \wedge dQ\wedge d\phi\nonumber\\
	&= -\frac{\lambda}{4\pi^2} \int_{u-{\rm plane}}dP\wedge dQ\nonumber\\
	&= -\frac{\lambda}{4\pi^2} \int_{u-{\rm plane}}d (P\wedge dQ)\nonumber\\
	&= -\frac{\lambda}{4\pi^2} \int_{\partial(u-{\rm plane})} P\wedge dQ\nonumber\\
	&= -\frac{\lambda}{4\pi^2} \sum_iP_i\int_{I_i} dQ\nonumber\\
	&= -\frac{\lambda}{4\pi^2} \sum_{i\geq 2}
			\left(\sum_{j\leq i-1}\frac{2\pi}{\sqrt{\lambda}}m_j\right)
			\left(\frac{2\pi}{\sqrt{\lambda}}n_i\right)\nonumber\\
	&= -\sum_{i\geq 2} \left(\sum_{j\leq i-1}m_j\right)n_i\label{SUMMjNi}\nonumber\\
	&= -\left(\sum_{j\leq 1}m_j\right)n_2- \left(\sum_{j\leq 2}m_j\right)n_3 \cdots. 
\end{align}
Here $\mathcal{M}_4$ is a 4-cycle coordinated by $u^1,u^2,\psi,\phi$.
In the 5th line we used the fact that $P$ is a constant at each $I_i$. Then in the next line the integral can be rewritten by \eqref{Defni} and the potential functions $P_i$ can be translated by adding a constant to all $P_i$. Using this ambiguity we set $P_1=0$. The first term of the final expression \eqref{SUMMjNi} ($i=2$) is equal to the number of the boxes in the lowest set of columns of the corresponding Young diagram. The second term is equal to the number of the boxes in the second lowest set of columns, and so forth (figure \ref{fig:YoungD}).

From the above calculations \eqref{CalcD3number}, \eqref{SUMMjNi}, we see a correspondence between the brane configuration and the number of the boxes in the Young diagram.  Namely, $k$, the number of the D3-branes ending on the D5-brane, corresponds to the vertical length of the Young diagram, and  $k'$, the number of the D1-branes embedded on the D5-brane, is the total number of the boxes in the Young diagram.  These are consistent with our conjectured relation.

\section{Conclusion}\label{Conclusion}
In this paper we propose the relation between the Young diagram and the brane configuration, and find the method to determine the brane configuration from the Young diagram. This relates the shape of the Young diagram, described by integers \eqref{Defni} and \eqref{Defmi},  and the brane configuration as follows.
These numbers are the number of boxes in the Young diagram. The correspondence is that $k$, the number of the D3-branes ending on the D5-brane, corresponds to the vertical length of the Young diagram \eqref{CalcD3number} and $k'$, the number of the D1-branes  embedded on the D5-brane corresponds to the total number of the boxes in the Young diagram. 
Once we are given a certain 't Hooft operator, we obtain a Young digram describing that operator. This information about the Young diagram directly requires the boundary condition via $n_i$ and $m_j$ \eqref{Defni} and \eqref{Defmi}. Then we relate this operator to a brane configuration according to eqs.~\eqref{Latestform1}-\eqref{Latestform4}.

We can propose some interesting future works. First, we can try to confirm this correspondence by a concrete calculation as in section \ref{SpecialCase} for the simplest case. 

Second, it is also an interesting future work to calculate physical quantities, such as expectation values of these 't Hooft operators and correlation functions with other operators in the string theory side and the gauge theory side.  In the gauge theory side, we can make use of localization technique\cite{Pestun:2007rz,Gomis:2011pf}.  In the string theory side we can compute these quantities from the classical action of the D5-brane.  It will be very interesting to compare these two results and check the AdS/CFT correspondence.

Finally, another interesting application is to consider a deformed 't Hooft operator. In this paper we only consider the simplest path --- straight line for the 't Hooft operator. When this path is deformed into a knotted configuration, the brane configuration becomes much more complicated. This topic is related to the knot homology as recently studied in \cite{Witten:2011zz,Gaiotto:2011nm}.

\section*{Acknowledgments}
We would like to thank Toshiaki Fujimori, Kazuo Hosomichi, Hiroaki Nakajima, Kazunobu Maruyoshi, Noriaki Ogawa, Shuichi Yokoyama for discussions and comments. 
We also thank the Yukawa Institute for Theoretical Physics at Kyoto University. 
Discussions during the YITP workshop on ``Field Theory and String Theory'' (YITP-W-13-12) 
were useful to complete this work. 
The work of K.N. was supported in part by the JSPS Research Fellowship for Young Scientists.
The work of S.Y. was supported in part by JSPS KAKENHI Grant No. 22740165.
\appendix
\section{SUSY in bulk space}\label{BulkSUSY}

We investigate supersymmetry in $AdS_5\times S^5$ spacetime with metric
\begin{equation}
ds^2=\frac{1}{y^2}(-dt^2+dy^2+dr^2+r^2d\psi^2+dx_3^2)+d\theta^2+\sin^2\theta d\phi^2.
\end{equation}
In order to preserve supersymmetry, the gravitino transformation must give zero,
\begin{align}\label{EpsilonEq}
\nabla_M\epsilon +\frac{i}{2^4}\Gamma^{M_1M_2\cdots M_5}F^{(5)}_{M_1M_2\cdots M_5}\Gamma_M\epsilon =0,\\
\nabla_M=\partial_M+\frac{1}{4}\Omega_M{}^{AB}\Gamma_{AB},
\end{align}
where gamma matrices with indices $M=t,r,\psi,x_3,y,\theta,\phi$ are $\Gamma_{t}:=E^A_t\Gamma_A=\frac{1}{y}\Gamma_0$ and so on. $\Gamma_{A},\ A=0,\dots,9$, are constant gamma matrices in 10-dimensional spacetime. They satisfy $\{\Gamma_A,\Gamma_B\}=2\eta_{AB}$ where $\eta_{AB}=\mathrm{diag}(-1,+1,\dots,+1)$.  We use the notation for antisymmetrized products of gamma matrices as
\begin{equation}
\Gamma_{A_1A_2\dots A_n}:=\frac{1}{n!} \sum_{\sigma\in \mathfrak{S}_n} \mathrm{sign}(\sigma)
\Gamma_{A_{\sigma(1)}}
\Gamma_{A_{\sigma(2)}}\cdots
\Gamma_{A_{\sigma(n)}}.
\end{equation}
 The SUSY parameter $\epsilon$ is a complex Weyl spinor which satisfies $\Gamma^{01\dots 9}\epsilon=\epsilon$.  In this paper we choose vielbein as 
\begin{align}
& E^0=\frac{dt}{y},\:\: E^1=\frac{dr}{y},\:\: E^2=\frac{rd\psi}{y},\:\: E^3=\frac{dx_3}{y},\:\: E^4=\frac{dy}{y},\nonumber\\
& E^5=d\theta,\:\: E^6=\sin\theta d\phi.
\end{align}
The spin connections $\Omega^{AB}=\Omega_M{}^{AB}E^M$ are related to vielbein as $dE^A=-\Omega^A_{\:\: B}E^B$, and calculated using this relation as follows.
\begin{align}
& \Omega^{04}=-\frac{dt}{y},\:\: \Omega^{12}=-d\psi,\:\: \Omega^{14}=-\frac{dr}{y},\nonumber\\
& \Omega^{24}=-\frac{rd\psi}{y}, \:\: \Omega^{34}=-\frac{dx_3}{y},\:\: \Omega^{56}=-\cos\theta d\phi,
\end{align}
and the other components are zero.
The equations \eqref{EpsilonEq} for 7 components, $M=t, r, \psi, x_3, y, \theta, \phi$, are 
\begin{subequations}
\renewcommand{\theequation}{\theparentequation--\roman{equation}}
\begin{align}
& \partial_t \epsilon -\frac{1+\gamma}{2y}\Gamma_{04}\epsilon =0\label{tcompEq},\\
& \partial_r \epsilon -\frac{1+\gamma}{2y}\Gamma_{14}\epsilon =0,\\
& \partial_{\psi} \epsilon -\frac{1}{2}\Gamma_{12}\epsilon -\frac{1+\gamma}{2y}\Gamma_{24}\epsilon =0,\\
& \partial_{x_3} \epsilon -\frac{1+\gamma}{2y}\Gamma_{34}\epsilon=0,\\
& \partial_y \epsilon +\frac{1}{2y}\gamma\epsilon=0,\\
& \partial_\theta \epsilon +\frac{1}{2}\gamma\Gamma_{45}\epsilon=0,\\
& \partial_\phi \epsilon -\frac{1}{2}e^{-\gamma\Gamma_{45}}\Gamma_{56}\epsilon=0,\label{phicompEq}
\end{align}
\end{subequations}
where we used the matrix $\gamma:=-i\Gamma_{0123}$.
Solving the equations \eqref{tcompEq}-\eqref{phicompEq} we obtain the supersymmetry parameter in the bulk spacetime.
\begin{equation}
\epsilon=e^{-\frac{\theta}{2}\gamma\Gamma_{45}}e^{\frac{\phi}{2}\Gamma_{56}}e^{-\frac{1}{2}\ln y\cdot \gamma}
e^{r\frac{1+\gamma}{2}\Gamma_{14}}
e^{x_3\frac{1+\gamma}{2}\Gamma_{34}}
e^{t\frac{1+\gamma}{2}\Gamma_{04}}
e^{\frac{\psi}{2}\Gamma_{12}}\epsilon_0,
\end{equation}
where $\epsilon_0$ is an arbitrary constant complex Weyl spinor.
For convenience, we define $\xi:=e^{\frac{\phi}{2}\Gamma_{56}}e^{\frac{\psi}{2}\Gamma_{12}}\epsilon_0$.  Then $\epsilon$ is rewritten as
\begin{equation}\label{SUSYParameterXi}
\epsilon=e^{-\frac{\theta}{2}\gamma\Gamma_{45}}e^{-\frac{1}{2}\ln y\cdot \gamma}
e^{r\frac{1+\gamma}{2}\Gamma_{14}}
e^{x_3\frac{1+\gamma}{2}\Gamma_{34}}
e^{t\frac{1+\gamma}{2}\Gamma_{04}}
\xi.
\end{equation}

\section{Kappa symmetry projection for D5-branes and D1-branes}\label{GammaForD5D1}

The kappa symmetry projection \cite{Cederwall:1996pv,Aganagic:1996pe,Cederwall:1996ri,Bergshoeff:1996tu,Aganagic:1996nn,Bergshoeff:1997kr,Skenderis:2002vf} plays a crucial role in our research. The supersymmetry with the parameters \eqref{SUSYParameterXi} which satisfy
\begin{equation}\label{GammaEpsilonEpsilon}
\Gamma\epsilon=\epsilon
\end{equation}
survives in the presence of a D-brane.   
Here the projection operator $\Gamma$ is defined for a D$p$-brane in type IIB string theory as
\begin{align}
& d^{p+1}\xi \cdot\Gamma := \left(-e^{-\Phi}(-\det(G_{\rm ind}+\mathcal{F}))^{-1/2}e^{\mathcal{F}}\chi\right)\Big|_{(p+1)-{\rm form}} \label{DefGamma},\\
& \chi := \sum_{n} \frac{1}{(2n)!} \hat{E}^{a_{2n}} \cdots \hat{E}^{a_1} \Gamma_{a_1\cdots a_s} K^n(-i) \label{DefChi},
\end{align}
where $\xi^i$,$i=0,\cdots, p$, are worldvolume coordinates, $\Phi$ is the dilaton which is zero now, $G_{\rm ind}$ is the induced metric of the D$p$-brane and $\hat{E}^A$ is the pull back of $E^A$ defined as $\hat{E}^A:= E^A_M \frac{\partial X^M}{\partial \xi^i} d\xi^i$. 

In this section we calculate the two cases of them--- a D$5$-brane and a D$1$-brane.
We now consider the situation in $AdS_5\times S^5$ spacetime formed by multiple D$3$-branes with metric
\begin{equation}
ds^2 = \frac{1}{y^2}(-dt^2+dy^2+dr^2+r^2d\psi^2+dx_3^2) + d\theta^2+\sin^2\theta d\phi^2,
\end{equation}
where we concentrate on the $S^2$ part of the $S^5$ and the AdS radius is set to unity.

\subsection{D5-brane}
First, let us consider the D5-brane with ansatz \cite{Karch:2000gx,Skenderis:2002vf}
\begin{equation}
x_3=\kappa y,\:\:\: \mathcal{F}=f\sin\theta \:d\theta\wedge d\phi,\:\:\: (\kappa,f: {\rm constant}).
\end{equation}
The induced metric of the D$5$-brane with coordinates $(t,r,\psi,y=\frac{1}{\kappa}x_3,\theta,\phi)$ is 
\begin{equation}
ds_{\rm D5}^2=\frac{1}{y^2}(-dt^2+dr^2+r^2d\psi^2+(\kappa^2+1)dy^2)+d\theta^2+\sin^2\theta d\phi^2.
\end{equation}
We need to calculate the determinant of
\begin{equation}
G_{\rm ind}+\mathcal{F}=
\left[\begin{array}{cccccc}
-1/y^2 &  &  &  &  &  \\
 & 1/y^2 &  &  &  &  \\
  &  & r^2/y^2 &  &  &  \\
   &  &  & (1+\frac{1}{\kappa^2})/y^2 &  &  \\
    &  &  &  & 1 & f\sin\theta \\
     &  &  &  & -f\sin\theta & \sin^2\theta\end{array}\right],
\end{equation}
where all empty components denote zeros.
The result is 
\begin{equation}
\sqrt{-\det(G_{\rm ind}+\mathcal{F})}=\frac{r\sin\theta}{y^4}\sqrt{1+1/\kappa^2}\sqrt{1+f^2}.
\end{equation}
Since $e^{\mathcal{F}}=1+f\sin^2\theta d\theta\wedge d\phi$,
\begin{align}
d^6\xi\cdot\Gamma_{\rm D5}
& =\left(-\frac{1}{\sqrt{1+1/\kappa^2}\sqrt{1+f^2}}\frac{y^4}{r\sin\theta}\Big(1+f\sin\theta d\theta\wedge d\phi\Big)\chi\right)\Big|_{\rm 6-form}\nonumber\\
& =-\frac{1}{\sqrt{1+1/\kappa^2}\sqrt{1+f^2}}\frac{y^4}{r\sin\theta}\Big(\chi\Big|_{\rm 6-form}+f\sin\theta d\theta\wedge d\phi\cdot \chi\Big|_{\rm 4-form}\Big).
\end{align}
$\chi|_{\rm 6-form}$ and $\chi|_{\rm 4-form}$ are 
\begin{align}
\chi|_{\rm 6-form} &= dt\:dr\:d\psi\: dy\: d\theta\: d\phi Ki\frac{r\sin\theta}{y^4}\left(\Gamma_{012356}+\frac{1}{\kappa}\Gamma_{012456}\right), \\
\chi|_{\rm 4-form} &= dt\:dr\:d\psi\: dy (-i)\frac{r}{y^4}\left(\Gamma_{0123}+\frac{1}{\kappa}\Gamma_{0124}\right).
\end{align}
We obtain the following result by putting together them.
\begin{equation}\label{GammaForD5p1}
\Gamma_{\rm D5}=\frac{-1}{\sqrt{(\kappa^2+1)(f^2+1)}}\gamma(K\Gamma_{56}+f)(\Gamma_{34}+\kappa).
\end{equation}
The necessary and sufficient condition for $\epsilon$ to satisfy $\Gamma_{\rm D5}\epsilon=\epsilon$ is $\kappa=-f$ and
\begin{equation}\label{KappaConditionD5}
(K\Gamma_{3456}+\gamma)\xi=0.
\end{equation}

\subsection{D1-brane}
Next, let us calculate the D1-brane case. The induced metric for the D1-brane with worldvolume coordinates $(t,y)$ is 
\begin{equation}
ds_{\rm D1}^2 = \frac{1}{y^2}(-dt^2+dr^2).
\end{equation}
Since the dilaton $\Phi$ is zero and there is no flux, $\mathcal{F}=0$,
\begin{equation}
d^2\xi\cdot \Gamma = -y^2\chi|_{\rm 2-form}.
\end{equation}
Substituting
\begin{equation}
\chi|_{\rm 2-form} = -dt\:dr K(-i)\frac{1}{y^2}\Gamma_{04},
\end{equation}
we obtain 
\begin{equation}
\Gamma_{\rm D1} = \Gamma_{04}K(-i).
\end{equation}
The necessary and sufficient condition for satisfying $\Gamma_{D1}\epsilon=\epsilon$ is
\begin{equation}\label{KappaConditionD1}
(iK\Gamma_{04}-1)\xi =0.
\end{equation}
Both the conditions \eqref{KappaConditionD5} and \eqref{KappaConditionD1} are satisfied in our bound state of a D5-brane and D1-branes.
\section{Derivation of $\Gamma$}\label{CalcGamma}
We calculate $\Gamma$ defined in \eqref{DefGamma} and \eqref{DefChi} for a D5-brane with worldvolume coordinates $(t,\psi,\phi,y,u^1,u^2)$. There is a flux on the D5-brane,
\begin{equation}
\mathcal{F}=dP(u)\wedge d\psi +dQ(u)\wedge d\phi.
\end{equation}
In our situation, the dilation is zero, and 
\begin{align}
e^\mathcal{F} &=1+\partial_aP du^a \wedge d\psi +\partial_a Q du^a \wedge d\phi - \partial_aP\partial_bQ\epsilon^{ab}d\psi\wedge  d\phi\wedge  du^1\wedge du^2,\\
(e^\mathcal{F}\chi)|_{\rm 6-from}
		&= \chi|_{\rm 6-form} +\chi|_{\rm 4-form}\cdot \partial_aPdu^a\wedge d\psi +\chi|_{\rm 4-form}\cdot \partial_aQdu^a\wedge  d\phi\nonumber\\
		&\hspace{3cm}	+\chi|_{\rm 2-form}\cdot(-\partial_aP\:\partial_bQ\epsilon^{ab})d\psi\wedge  d\phi\wedge  du^1\wedge du^2 \label{eFX}. 
\end{align}
Here the first $\chi|_{\rm 4-form}$ in the expression \eqref{eFX} is proportional to $dt\wedge d\phi\wedge dy\wedge du^b, (b\neq a)$, while the second is proportional to $dt\wedge d\psi\wedge  dy\wedge  du^b, (b\neq a)$ and we use the notation 
\begin{equation}
\{A,B\}:=\epsilon^{ab}\partial_aA\partial_bB=\frac{\partial A}{\partial u^1}\frac{\partial B}{\partial u^2}-\frac{\partial A}{\partial u^2}\frac{\partial B}{\partial u^1},
\end{equation}
in the following.
Each term of eq.\eqref{eFX} is calculated as follows.
\begin{subequations}
\renewcommand{\theequation}{\theparentequation--\roman{equation}}
\begin{align}
&\chi|_{\rm 6-form} 
	= d^6\xi\cdot \frac{s\sin\theta}{y^2}\Big(\{z,\theta\}\Gamma_{35}+\{s,\theta\}\Gamma_{15}-s^2\{\frac{z}{s},\theta\}\Gamma_{1345}+\{s,z\}\Gamma_{13}\Big)\Gamma_{04}\Gamma_{62}K(-i),\\
&\chi|_{\rm 4-form} \cdot \partial_aP du^ad\psi
	= \frac{\sin\theta}{y^2}\Big(s^2\{P,\frac{z}{s}\}\Gamma_{13}-\{P,s\}\Gamma_{14}-\{P,z\}\Gamma_{34}\nonumber\\
	&\hspace{3cm}	+s\{P,\theta\}\Gamma_{15}+z\{P,\theta\}\Gamma_{35}+\{P,\theta\}\Gamma_{45}\Big)\Gamma_{60}(-i)d^6\xi,\\
&\chi|_{\rm 4-form} \cdot \partial_aQ du^ad\psi
	= \frac{s}{y^2}\Big(-s^2\{Q,\frac{z}{s}\}\Gamma_{13}+\{Q,s\}\Gamma_{14}+\{Q,z\}\Gamma_{34}\nonumber\\
	&\hspace{3cm}	-s\{Q,\theta\}\Gamma_{15}-z\{Q,\theta\}\Gamma_{35}-\{Q,\theta\}\Gamma_{45}\Big)\Gamma_{20}(-i)d^6\xi,\\
&\chi|_{\rm 2-form}\cdot(-\partial_aP\partial_bQ\epsilon^{ab})d\psi\: d\phi\: dt\: dy
	= \frac{1}{y^2}(\epsilon^{ab}\partial_aP\partial_bQ)(s\Gamma_{14}+z\Gamma_{34}+1)\Gamma_{04}K(-i)\cdot d^6\xi,
\end{align}
\end{subequations}
where $d^6\xi =dt\wedge  d\psi\wedge  d\phi\wedge  dy\wedge  du^1\wedge  du^2$.

In the definition \eqref{DefGamma}, $\mathcal{L}_{\rm DBI}$ is
\begin{equation}
\mathcal{L}_{\rm DBI} =\sqrt{-\det(G_{\rm ind}+\mathcal{F})}=:\frac{W}{y^2}.
\end{equation}
Under our ansatz, see eq.\eqref{Ansatz} in section \ref{OurAnsatz}, the induced metric $G_{\rm ind}$ is 
\begin{align}
ds_{\rm ind}^2=-\frac{1}{y^2}dt^2+s^2d\psi^2+\sin^2\theta d\phi^2 + \frac{\beta}{y^2}dy^2
	+h_{ij} du^i du^j +\frac{\partial_a\beta}{y}du^ady,\\
h_{ij}:=\sum_{\lambda=s,z,\theta}\partial_i \lambda \partial_j \lambda.
\end{align}
We define a convenient variable $\beta :=1+s^2+z^2$.
$W$ is calculated as the following determinant.
\begin{align}
W^2
&=-y^4\det
\left[\begin{array}{c|c|c|c}
-1/y^2 &  &  &  \\
\hline
  & \left.\begin{array}{cc}s^2 &  \\ & \sin^2\theta\end{array}\right.  &  & \left.\begin{array}{cc}-J_1 & -J_2 \\-L_1 & -L_2\end{array}\right. \\
\hline  &   & \frac{\beta}{y^2} & \left.\begin{array}{cc}\frac{1}{2y}\partial_1\beta & \frac{1}{2y}\partial_2\beta\end{array}\right. \\
\hline  &  \left.\begin{array}{cc}J_1 & L_1 \\J_2 & L_2\end{array}\right. & \left.\begin{array}{c}\frac{1}{2y}\partial_1\beta \\\frac{1}{2y}\partial_2\beta\end{array}\right. & \left.\begin{array}{cc}h_{11} & h_{12} \\h_{21} & h_{22}\end{array}\right.\end{array}\right]\nonumber\\
&=\det
\left[\begin{array}{c|c|c}
   \left.\begin{array}{cc}s^2 &  \\ & \sin^2\theta\end{array}\right.  &  & \left.\begin{array}{cc}-J_1 & -J_2 \\-L_1 & -L_2\end{array}\right. \\
\hline  
  & {\beta} & \left.\begin{array}{cc}\frac{1}{2}\partial_1\beta & \frac{1}{2}\partial_2\beta\end{array}\right. \\
\hline   
 \left.\begin{array}{cc}J_1 & L_1 \\J_2 & L_2\end{array}\right. & \left.\begin{array}{c}\frac{1}{2}\partial_1\beta \\\frac{1}{2}\partial_2\beta\end{array}\right. & \left.\begin{array}{cc}h_{11} & h_{12} \\h_{21} & h_{22}\end{array}\right.\end{array}\right],
 \end{align}
where $J_a:=\partial P/\partial u^a$ and $L_a:=\partial Q/\partial u^a$.
To calculate this determinant the following formula is convenient.
\begin{equation}
\det\left[\begin{array}{cc}A & D \\C & B\end{array}\right]
=\det A\cdot \det(B-CA^{-1}D).
\end{equation}
We use this formula for
\begin{align*}
& A=\left[\begin{array}{ccc}s^2 &  &  \\ & \sin^2\theta &  \\ &  & \beta\end{array}\right],\:\:
D=\left[\begin{array}{cc}-J_1 & -J_2 \\-L_1 & -L_2 \\\frac{1}{2}\partial_1\beta & \frac{1}{2}\partial_2\beta\end{array}\right],\\
& C=\left[\begin{array}{ccc}J_1 & L_1 & \frac{1}{2}\partial_1 \beta \\J_2 & L_2 & \frac{1}{2}\partial_2\beta\end{array}\right],\:\:
B=\left[\begin{array}{cc}h_{11} & h_{12} \\h_{21} & h_{22}\end{array}\right].
\end{align*}
Then $W$ is written explicitly as
\begin{align}\label{DefW}
W^2=& s^2\sin^2\theta\{s,z\}^2\nonumber\\
		&+s^2\sin^2\theta ((z^2+1)\{s,\theta\}^2+(s^2+1)\{z,\theta\}^2-2sz\{s,\theta\}\{z,\theta\})\nonumber\\
		& +\sin^2\theta ((z^2+1)\{s,P\}^2+(s^2+1)\{z,P\}^2-2sz\{s,P\}\{z,P\})\nonumber\\
		& +s^2 ((z^2+1)\{s,Q\}^2+(s^2+1)\{z,Q\}^2-2sz\{s,Q\}\{z,Q\})\nonumber\\
		& +\beta \{P,Q\}^2.
\end{align}
Summarizing the above, the operator $\Gamma$ is 
\begin{equation}\label{Gamma}
\Gamma = \frac{1}{W} \Big\{
s\sin\theta \:\mathcal{A} \Gamma_{62}K(-i)\Gamma_{04}
+\sin\theta \:\mathcal{B} (-i)\Gamma_{60}
-s\:\mathcal{C}(-i)\Gamma_{20}
+\mathcal{D} K(-i) \Gamma_{04}
\Big\},
\end{equation}
where
\begin{subequations}
\renewcommand{\theequation}{\theparentequation--\roman{equation}}
\begin{align}
&\mathcal{A} := -\{s,z\}\Gamma_{13}-\{s,\theta\}\Gamma_{15} -\{z,\theta\}\Gamma_{35} + s^2\{\frac{z}{s},\theta\}\Gamma_{1345},\\ 
&\mathcal{B} := -\{P,\frac{z}{s}\}\Gamma_{13}+\{P,s\}\Gamma_{14}+ \{P,z\}\Gamma_{34} -s\{P,\theta\}\Gamma_{15} -z\{P,\theta\}\Gamma_{35}-\{P,\theta\}\Gamma_{45},\\
&\mathcal{C} := -\{Q,\frac{z}{s}\}\Gamma_{13}+\{Q,s\}\Gamma_{14}+ \{Q,z\}\Gamma_{34} -s\{Q,\theta\}\Gamma_{15} -z\{Q,\theta\}\Gamma_{35}-\{Q,\theta\}\Gamma_{45},\\
&\mathcal{D} := -\{P,Q\} (1+s\Gamma_{14}+z\Gamma_{34}),
\end{align}
\end{subequations}
$\mathcal{C}$ is obtained by replacing all $P$'s in $\mathcal{B}$ by $Q$'s, and $W$ is given by eq.\eqref{DefW}.

\providecommand{\href}[2]{#2}\begingroup\raggedright\endgroup

\end{document}